\documentclass[apsrev,reprint]{revtex4-2}
\usepackage{inputenc} 
\usepackage{amsmath} 
\usepackage{amssymb}
\usepackage{graphicx} 
\usepackage{nccmath}
\usepackage{caption} 
\usepackage{siunitx} 
\usepackage{color} 
\graphicspath{{figures/}} 
\usepackage{subcaption}
\usepackage{tabularx}
\usepackage{amsthm}
\usepackage{mathtools}
\usepackage{hyperref}
\usepackage[margin = 1in]{geometry}
\usepackage{natbib}
\newcommand{\figref}[1]{Fig. \ref{#1}}
\newcommand{\mybig}{\bBigg@{4}}

\makeatletter
\newcommand{\Biggg}{\bBigg@{3.5}}
\makeatother

\begin{document}
\begin{abstract}
Understanding the structural evolution of granular systems is a long-standing problem. A recently proposed theory for such dynamics in two dimensions predicts that steady states of very dense systems satisfy detailed-balance. 
We analyse analytically and numerically the steady states of this theory in systems of arbitrary density and report the following. 
1. We discover that all such dynamics almost certainly possess only one physical steady state, which may or may not satisfy detailed balance. 
2. We show rigorously that, if a detailed balance solution is possible then it is unique. The above two results correct an erroneous conjecture in the literature. 
3. We show rigorously that the detailed-balance solutions in very dense systems are globally stable, extending the local stability found for these solutions in the literature. 
4. In view of recent experimental observations of robust detailed balance steady states in very dilute cyclically sheared systems, our results point to a self-organisation of process rates in dynamic granular systems.
\end{abstract}

\title{Steady states of two-dimensional granular systems are unique, stable, \\
and sometimes satisfy detailed balance}
\author{Alex D. C. Myhill}
    \affiliation{Cavendish Laboratory, University of Cambridge, JJ Thomson Avenue, Cambridge, CB3 0HE, UK}
    \altaffiliation{Bullard Laboratory, University of Cambridge, Madingley Road, Cambridge CB3 0EZ, UK}
\author{Raphael Blumenfeld}\email{rbb11@cam.ac.uk}
    \altaffiliation{Gonville \& Caius College, University of Cambridge,
    Trinity Street, Cambridge CB2 1TA, UK}
    \affiliation{ESE, Imperial College London,
    Prince Consort Rd, London SW7 2AZ, UK}
\maketitle

{\bf 1. Introduction}
Granular matter is ubiquitous in nature and plays a major role in our everyday life. Its near-indifference to thermal fluctuations has earned it a recognition as a new form of matter~\cite{duran2012sands}. In spite of many decades of intensive theoretical, numerical, and experimental investigations into this form of matter, new aspects of its rich and complex behaviour are being discovered. 
The sensitivity of the large-scale behaviour and properties to the particle scale characteristics and structure has hindered the modelling of granular matter to date~\cite{vogel_moving_2003,edwards_transmission_1989}. Consequently, of key significance is the modelling of granular dynamic evolution and the mechanically stable structures that such dynamics settle into. In particular, when the dynamics are quasistatic, the steady-state dynamics determine those stable structures and it is on this type of dynamics that we focus here.

\begin{figure} [h]
\centering
\includegraphics[width=\columnwidth]{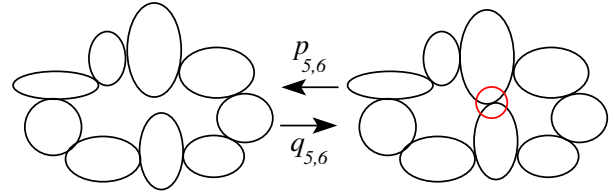}
 \caption{A sketch of cellular splitting and merging events: when the circled contact breaks, the two cells merge, $5 + 6\to 9$ and when it is made the higher order cell splits, $9\to 5 + 6$.}
\label{CellTransition}
\end{figure}

Several methods have been proposed to describe and model the evolution of the underlying structure during quasistatic dynamics~\cite{Meetal09,Ge10,Zietal11,PuTo20,Deetal22}. A general way to describe mechanically stable granular structures in two-dimensions (2D) is based on what is known as the cell order distribution (COD)~\cite{matsushima_universal_2014,matsushima_fundamental_2017,sun_friction-controlled_2020}. A cell is the smallest void (loop) in the system, surrounded by particles in contact and its order is defined as the number of particles in contact surrounding it. During quasi-static dynamics, the COD evolves by intergranular contacts being made and broken, which split and merge cells, respectively. Such a process is shown schematically in Fig~\ref{CellTransition}.
The dynamic equations for the COD evolution are~\cite{wanjura_structural_2020}: 
\begin{align}
  & \dot{Q_k} = \frac{1}{2}\sum_{i=3}^{k-1} \left(p_{i,j}Q_iQ_j-q_{i,j}Q_{i+j-2}\right)\left(1 + \delta_{i,k-i+2}\right) - \nonumber \\
   & \sum_{i=k+1}^{\mathcal{C}} \left(p_{k,j}Q_kQ_{i-k+2}-q_{k,{i-k+2}}Q_{i-k+2}\right)\left(1+ \delta_{i,2k-2}\right) \nonumber \\
   & + Q_k \hspace{-1em}\sum_{\text{all possible}\atop\text{processes $i,j$}} \left(p_{i,j}Q_iQ_j-q_{i,j}Q_{i+j-2}\right) \ .
\label{eqn:master}
\end{align}
In these equations, $Q_k$ is the fraction of cells of order $k$, referred to in the following as $k$-cells. $p_{i,j}$ is the merging rate of $i$- and $j$-cells into an $(i+j-2)$-cell, $q_{i,j}$ is the rate of the splitting of an $(i+j-2)$-cell into an $i$- and a $j$-cell, and $\mathcal{C}$ is the highest possible cell order in the system. The factor $1/2$ and the $\delta$-functions ensure correct counting. The last term on the right hand side is needed because the total number of cells changes with each merging or splitting event, which changes the fractions $Q_k$. Rattlers, which are particles with one or no contact, were ignored in these equations, which is a good approximation for dense systems with low-order cells. Including rattlers in the analyses that follow is possible, but the added complication does not add much insight and we disregard them. 

\textcite{wanjura_structural_2020} found that, under some conditions, the steady-state cell order transitions of these far-from-equilibrium systems satisfy detailed balance, when the cell orders do not exceed six, a result corrected later to five~\cite{wanjura_correction}. The steady states of the evolution equations were also shown to be locally stable.
Recent experiments on quasi-statically sheared 2D granular systems~\cite{sun_experimental_2021} have revealed a surprising observation -- they always settled into steady states that satisfy detailed balance. Such robustness suggest that these steady states are not only stable, but also may be unavoidable. Moreover, these observations appear to contradict the paradigm that steady states of non-equilibrium dynamics cannot satisfy detailed balance~\cite{klein_principle_1955}. 
Motivated by these experimental observations, we analyse here eqs.~\eqref{eqn:master} in detail. We investigate the conditions for detailed balance and the properties of such steady states. We show that: 
(i) if a steady state satisfies detailed balance in systems where $\mathcal{C}=6,7$ then it is the only possible steady state; 
(ii) there is strong numerical evidence that, in systems where $\mathcal{C}=6,7$, or $8$, only one physical steady state is possible, whether or not it satisfies detailed balance;
(iii) the steady state in systems where $\mathcal{C}=5$ is not only locally but also globally stable.
These findings provide a partial explanation for the observed convergence to detailed balanced steady states  in~\cite{sun_experimental_2021}. \\

\section{Analysis of the steady state solutions} 
\subsection{General steady states}
The back-and-forth processes $(i) + (j)\rightleftharpoons (i+j-2)$ are equivalent to chemical reactions in a multi-component reactive system. Their net flux is $\eta_{i,j}=p_{i,j}Q_iQ_j-q_{i,j}Q_{i+j-2}$ and $\eta_{i,j}=0$ when they are balanced. At steady state, the sum of all the processes, $\sum_{i,j}\bar{\eta}_{i,j}$, vanishes by definition and eqs.~\eqref{eqn:master} reduce to 
\begin{align}
	0 & = \frac{1}{2} \sum_{i = 3}^{k - 1} \bar{\eta}_{i,k-i+2} (1 + \delta_{i,k-i+2}) \nonumber \\
	& \phantom{{} = } - \sum_{i = k + 1}^{C} \bar{\eta}_{k,i-k+2}(1+ \delta_{i,2k-2}) \ ,
	\label{eqn:ss}
\end{align}
in which the bars indicate steady state values. 

Given $\mathcal{C}$, there can be $(\mathcal{C}-2)^2/4$ or $\left[(\mathcal{C}-2)^2-1\right]/4$ processes, when $\mathcal{C}$ is even or odd, respectively. Focusing on the even case, for brevity, it is useful to rewrite eqs. \eqref{eqn:ss} as 
\begin{equation}
H \cdot \bar{\boldsymbol \eta}= 0 \ .
\label{Aeta}
\end{equation}
Here, $H$ is a $(\mathcal{C}-2)\times\left[(\mathcal{C}-2)^2/4\right]$ matrix and the vector $\bar{\boldsymbol \eta}$'s components are all the steady-state $\eta$-fluxes. Thus, $\bar{\boldsymbol\eta}$ must exist within the null space of $H$.  
The normalisation constraint, $\sum_k \bar{Q}_k = 1$, reduces the number of independent first-order equations in (\ref{Aeta}) to $\mathcal{C}-3$. 
Thus, by B\'{e}zout's theorem~\cite{vogel1984lectures} and, since $\eta_{i,j}$ are quadratic in the $Q_i$s, the maximal number of solutions is $2^{\mathcal{C}-3}$ for any given set of rates, $p_{i,j}$ and $q_{i,j}$.

Below, we show analytically that, at least up to $\mathcal{C}=7$, only one of these solutions is physical -- the detailed balance steady state (when it exists), in which $\eta_{i,j}=0$ for all $\ i,j$. 
Indeed, an extensive numerical investigation over a wide range of parameters supports this conclusion -- all other numerical solutions included either imaginary or negative $Q_k$ fractions.

\subsection{The detailed-balance steady state is unique} 
\begin{figure} [h]
\centering
    \includegraphics[width=\columnwidth]{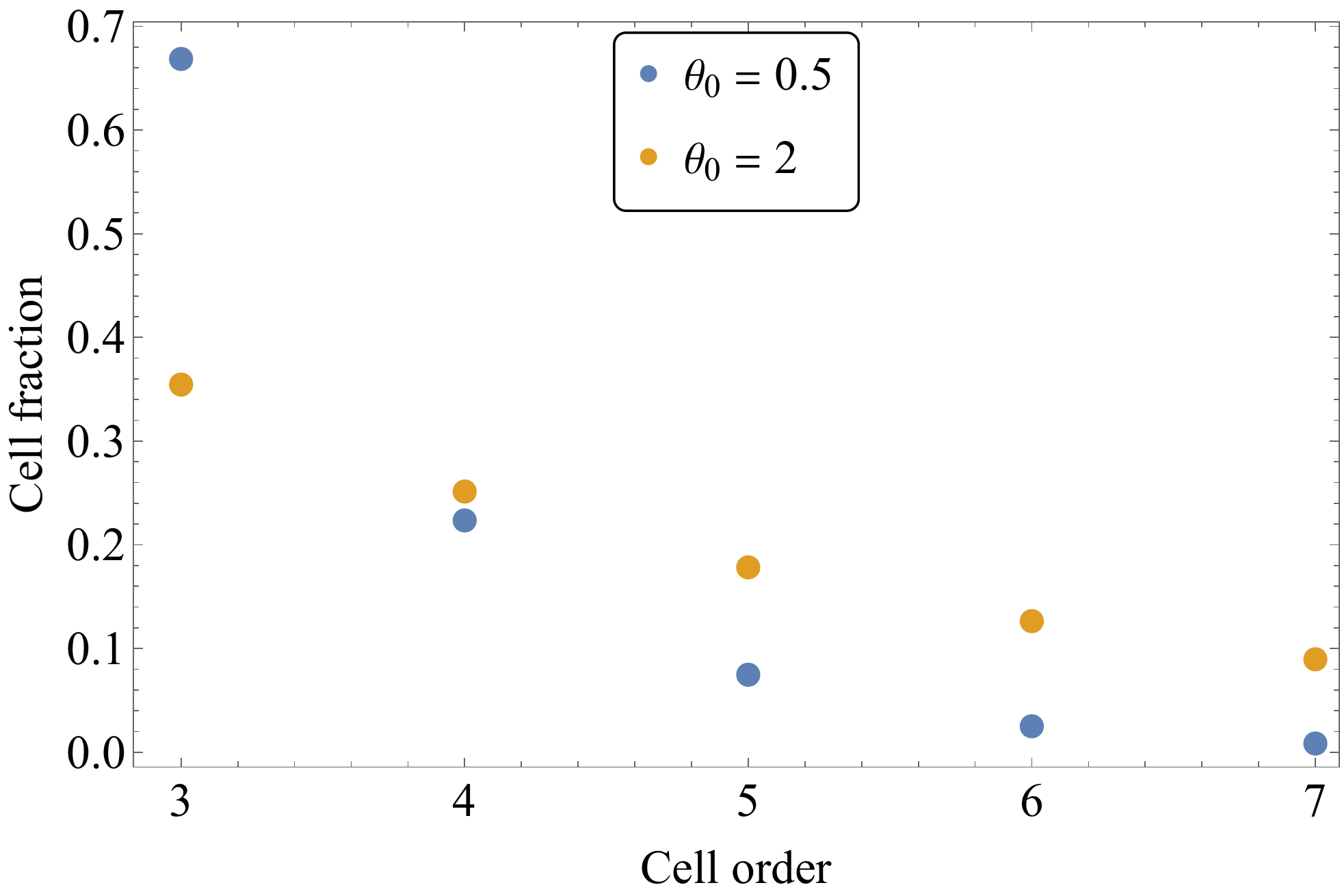}
    \caption{Two examples of detailed-balance steady states, for which $\mathcal{C} = 7$ and $\theta_{i,j} = \theta_0 = 0.5$ and $2$. }
\label{fig:ssexam}
\end{figure}

The uniqueness of the detailed balance steady state was established for $\mathcal{C}<6$~\cite{wanjura_structural_2020,wanjura_correction}. We now extend this result to systems comprising arbitrary orders. 
Defining $\theta_{i,j} = p_{i,j}/q_{i,j}$, each balanced process satisfies $\bar{Q}_{i + j - 2} = \theta_{i,j} \bar{Q}_i \bar{Q}_j$. It follows that
\begin{equation}
	\bar{Q}_k  = \bar{Q}_3^{k - 2} \prod_{i = 3}^{k - 1} \theta_{3,i} \ .
	\label{barQ}
\end{equation} 
The detailed-balance steady-state solution depends solely on the ratios $\theta_{i,j}$ and on $\bar{Q}_3$. The value of $\bar{Q}_3$ can then be found from the normalisation condition:
\begin{equation}
	 \bar{Q}_3 + \sum_{k = 2}^{C} \left\{\bar{Q}_3^{k} \prod_{i = 3}^{k+1} \theta_{3,i}\right\} = 1 \ .
	 \label{eq:qbarss}
\end{equation}
We note in passing that the values of $\theta_{i,j}$ are not independent because cells of order $k>5$ can be formed by more than one process~\cite{wanjura_thesis}. An example of a system in which detailed balance is possible (indeed observed) is the experimental steady states observed in~\cite{sun_friction-controlled_2020}, which satisfy $\theta_{i,j} = \theta_0$ for all $i,j$. This gives rise to an exponentially decaying COD. Imposing such a condition, we calculate explicitly the COD of the emerging detailed-balance steady state of two systems in which $\mathcal{C} = 7$ and $\theta_0=0.5$ and $2$. These are shown in~\figref{fig:ssexam}.

Nevertheless, this dependence does not affect our following argument regarding the uniqueness of the detailed balance steady state. Since the rates $p_{i,j}$ and $q_{i,j}$ must be non-negative then $\theta_{i,j}\geq0$, for all $i,j$, and the left hand side of \eqref{eq:qbarss} is a monotonically increasing function of $\bar{Q}_3$.  
 Additionally, for $\mathcal{C}>3$, it vanishes at $\bar{Q}_3=0$ and exceeds 1 at $\bar{Q}_3=1$. Thus, only one solution exists in the range $0<\bar{Q}_3<1$ and this is the only possible detailed-balance solution. This conclusion holds for both odd and even values of $C$. 
It follows that, if a system evolves into a detailed-balance steady state then that state is unique and independent of initial conditions. This goes some way to explain the robustness of the recently observed detailed-balance in \cite{sun_experimental_2021}.

\subsection{The case $\mathcal{C}=6$} \label{sec:dbss}
The next question is whether or not there also exist steady states that do not satisfy detailed balance. 
There are $\left[(\mathcal{C}-2)^2-1\right]/4$ or $(\mathcal{C}-2)^2/4$ potential finite values of the fluxes $\bar{\eta}_{i,j}$ to determine from eqs. (\ref{eqn:ss}), for $\mathcal{C}$ odd or even, respectively. With only $\mathcal{C}-3$ independent eqs.  available, in the case of $\mathcal{C} = 5$ there are two eqs. for two unknowns. It follows that detailed balance is the only steady state for systems up to $\mathcal{C}=5$. Since such systems have only two processes, $3+3\leftrightharpoons4$ and $3+4\leftrightharpoons5$, each must be balanced separately, as no cycle is possible~\cite{wanjura_correction}. This detailed balance is then straightforward. 
However, since $\bar{\eta}_{i,j}$ are underdetermined for $\mathcal{C}>5$, it was conjectured that, in addition to detailed-balance, such systems support an infinite number of other stable steady states~\cite{wanjura_structural_2020,wanjura_correction}. To investigate this conjecture, we re-examine the steady state.

Focusing initially on the case $\mathcal{C}=6$, there are four processes: $\eta_{3,3}$, $\eta_{3,4}$, $\eta_{3,5}$, and $\eta_{4,4}$, but only three independent equations. Rewriting (\ref{eqn:ss}) as
\begin{equation}
	\begin{bmatrix}
	-2 & -1 & -1 & 0 \\
	1 & -1 & 0 & -2 \\
	0 & 1 & -1 & 0\\
	0 & 0 & 1 & 1
	\end{bmatrix} 
	\begin{bmatrix}
	\eta_{3,3}\\
	\eta_{3,4}\\
	\eta_{3,5}\\
	\eta_{4,4}
	\end{bmatrix} = \mathbf{0} \ ,
\end{equation}
leaves one flux underdetermined. We parameterise the solution by $\eta_{3,3} \equiv A$:
\begin{equation}
	\boldsymbol{\eta} = \begin{bmatrix}
	\eta_{3,3}\\
	\eta_{3,4}\\
	\eta_{3,5}\\
	\eta_{4,4}
	\end{bmatrix}=\begin{bmatrix}
	A\\
	-A\\
	-A\\
	A
	\end{bmatrix} \ .
\label{etaalpha}
\end{equation}
It should be noted that, for any finite value of $A$, this steady state involves a cycle, $4 + 4\to6\to3 + 5\to6\to4 + 4$~\cite{klein_principle_1955}.
Using (\ref{barQ}) and (\ref{etaalpha}), the steady-state cell fractions are:
\begin{subequations}
\begin{align}
	\bar{Q}_4 & = (p_{3,3}\bar{Q}_3^2- A)/q_{3,3}, \label{eqn:101}\\
	\bar{Q}_5 & = (p_{3,4} \bar{Q}_3 \bar{Q}_4 + A)/q_{3,4}, \label{eqn:102} \\
	\bar{Q}_{6} & = (p_{3,5} \bar{Q}_3 \bar{Q}_5 + A)/q_{3,5}, \label{eqn:103}\\
	\bar{Q}_{6} & = (p_{4,4} \bar{Q}_4^2 - A)/q_{4,4} \ . \label{eqn:104}
\end{align} 
\label{etaalpha2}
\end{subequations} 
Eliminating variables, and imposing normalisation, yields a cumbersome equation for $\bar{Q}_3$, which in all but the simplest cases, can only be solved numerically. It is the dependence on the continuous parameter $A$ that led to the conjecture in \cite{wanjura_structural_2020} that there is an infinite family of solutions in these systems.

We show next that this is not the case, namely, if $A=0$ is a solution then no other solution exists. Firstly, note that, when $A=0$, 
$\theta_{i,j}Q_iQ_j=Q_{i+j-2}$. Eliminating $Q_4,Q_5$ and $Q_6$ from eqs. (\ref{etaalpha2}), we obtain $\theta_{3,3}\theta_{4,4} = \theta_{3,4} \theta_{3,5}$. \\
Since $p_{i,j},q_{i,j}\geq0$ then eqs. \eqref{eqn:103} and \eqref{eqn:104} imply that $\theta_{4,4}\bar{Q}_4^2 > \bar{Q}_{6}$ and $\bar{Q}_{6} > \theta_{3,5} \bar{Q}_3 \bar{Q}_5$ when $A > 0$. Additionally, from \eqref{eqn:102}, we have $\bar{Q}_5 > \theta_{3,4} \bar{Q}_3\bar{Q}_4$. Taken together, these yield $\theta_{4,4}\bar{Q}_4^2 > \theta_{3,4} \theta_{3,5} \bar{Q}_3^2 \bar{Q}_4$. Now, using the detailed balance condition and eliminating $\bar{Q}_6$, we have $\bar{Q}_4 > \theta_{3,3} \bar{Q}_3^2$, but \eqref{eqn:101} implies $\bar{Q}_4 < \theta_{3,3} \bar{Q}_3^2$. We have arrived at a contradiction, which means that $A>0$ cannot be a solution. A similar chain of analysis shows that $A<0$ is also impossible. It follows that, if the detailed balance solution, $A=0$, exists, then it is the only possible steady state. 

In the supplementary material, we use a similar analysis to show that the same conclusion holds for $\mathcal{C}=7$. This conclusion improves on the conjecture in ~\cite{wanjura_structural_2020}. We believe that this line of proof can be extended to $\mathcal{C}>7$, although not without substantial effort. However, a different approach is required for a general proof. 

These results, combined with the extensive numerical investigations reported below, and the experimental observations in \cite{sun_experimental_2021}, lead us to conjecture that, when a detailed balance steady state exists, it is {\it the only possible steady state} for any value of $\mathcal{C}$. \\

\subsection{Numerical investigation of non-detailed-balance steady states}
The rates parameter space is infinitely large and most combinations of rates lead to non-detailed-balance solutions (with detailed balance only possible if the relation established in \ref{sec:dbss} is satisfied, i.e. $\theta_{3,3}\theta_{4,4} = \theta_{3,4}\theta_{3,5}$). To understand the nature of these solutions, we explored the parameter space numerically (see also supplementary material). Starting with $\mathcal{C} = 6$, we tested the $4^8=65536$ rate value combinations when each rate can assume any of the four values: $0.1$, $0.5$, $1.0$, and $3.0$. For each combination, we found all the solutions and noted the number of solutions. Unexpectedly, each combination gave rise to only one physical solution, with all the others containing either negative or complex values of some $\bar{Q}_k$.

To test the potential generality of this surprising observation, we solved numerically for the steady-state solutions in systems where $\mathcal{C} = 7$ and $8$. These have, respectively, $6$ and $9$ processes and $12$ and $18$ variable rates. Owing to the required larger computational resources, we tested only $6$ rate combinations for $\mathcal{C} = 7$: $\{0.1,0.5\},\{0.1,1\},\{0.1,2\},\{0.5,1\},\{0.5,2\},\{1,2\}$, and in total $6\times 2^{12}=24576$ different systems. In each test, the values of $p_{i,j}$ and $q_{i,j}$ can take either of the pair of values noted. For $\mathcal{C} = 8$, we used the set of rates $\{0.1,0.5\}$ and $\{0.5,2\}$, and in total $2\times 2^{18}=524288$ different systems. We found that in none of these systems was there more than one physical steady state solution.

\begin{figure}
\centering
    \includegraphics[width=\columnwidth]{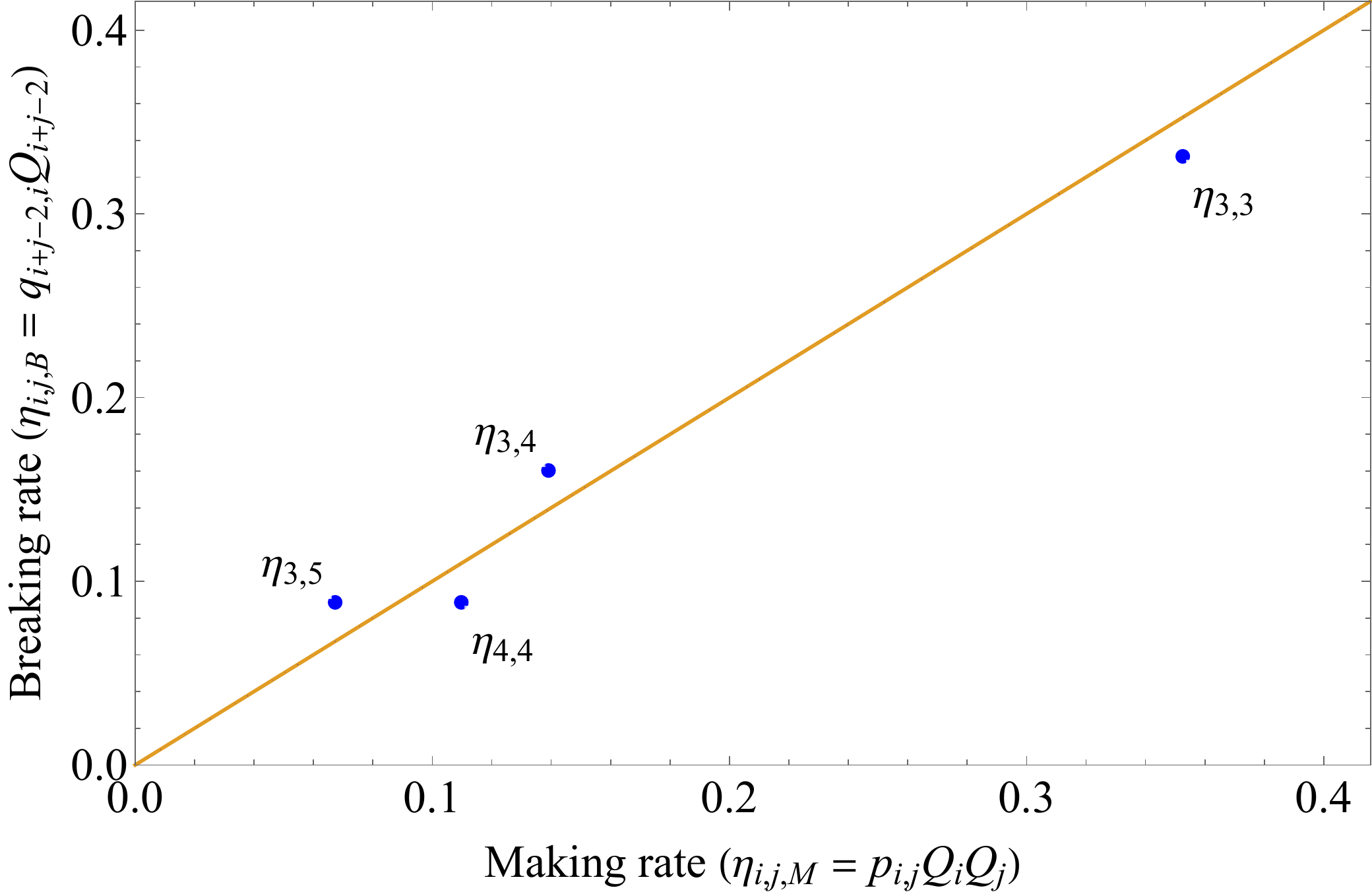}
    \caption{The unique steady state for $\mathcal{C} = 6$, with the parameters $p_{3,3} = 2$, $p_{i,j} = 1$ for all other $i,j$ and $q_{i,j} = 1$ for all $i,j$. The labels indicate the respective rates. The steady state rates do not lie on the solid line and, therefore, the system does not satisfy detailed-balance. }
\label{fig:nondb}
\end{figure}

Based on these investigations, we conjecture that, for any choice of constant rates, there is {\it only one physical solution} regardless of the upper order, $\mathcal{C}$. 
In Fig.~\ref{fig:nondb}, we show an example of the one non-detailed-balance solution when $\mathcal{C} = 6$, for a set of rate parameters that also admits five other non-physical solutions (listed in the supplemental material). It can be observed in Fig.~\ref{fig:nondb} that the difference between the breaking and making rates are the same for all rates. In particular, the vertical offset from the detailed balance line are the same and equal to $A$. A typical figure for $\mathcal{C} = 8$, for a system that also does not satisfy DB, is shown in the supplementary material.\\

\color{black}
\section{Global stability for $\mathcal{C}=5$} 
A linear analysis of the steady states of eqs. \eqref{eqn:ss} has shown them to be asymptotically stable~\cite{wanjura_structural_2020}, a prediction that has been supported experimentally~\cite{sun_experimental_2021}.
This, however, does not preclude possible limit cycles around the steady state away from the linear regime. We investigate next the global stability of the solution and show that, at least for the dense system comprising cells of orders $3-5$, no such cycles exist.

Using the normalisation condition to eliminate $Q_5$, the independent evolution equations can be written as
\begin{equation}
\begin{aligned}
\dot{Q}_3 & =(Q_3 - 2)\kappa_{3,3} + R(Q_3 - 1) \kappa_{3,4}, \\
\dot{Q}_4 & = (Q_4 + 1)\kappa_{3,3}+  R(Q_4 - 1) \kappa_{3,4} \ , 	
\end{aligned}
\label{eqn:468def}
\end{equation}
in which the rates $q_{i,j}$ and $\theta_{i,j}$ are assumed for now to be constant (more on this condition below), $R\equiv q_{3,4}/q_{3,3}$, $\kappa_{3,3}\equiv\theta_{3,3}Q_3^2-Q_4$, $\kappa_{3,4}\equiv\theta_{3,4}Q_3 Q_4+Q_3+Q_4-1$, and time is scaled: $t\to t'\equiv q_{3,3} t$, such that $\dot{Q}_k=dQ_k/dt'$. 

The fractions are constrained by $Q_3, Q_4 \geq 0$ and $Q_3 + Q_4 \leq 1$. We analyse eqs. (\ref{eqn:468def}) within this region of the $Q_3-Q_4$ plane, using the theorems of Bendixson and Poincar\'{e}-Bendixson~\cite{Perko_1996_Diff_Eqn}. The former states that, in two-variables dynamical systems, $\dot{x} = f(x,y)$ and $\dot{y} = g(x,y)$, if $V(x,y) = \partial_x(f) + \partial_y(g)$ is non-zero and has the same sign throughout a simply-connected $x-y$ domain, then no closed orbits can lie within that domain~\cite{Perko_1996_Diff_Eqn}. We define
\begin{align}
	V(Q_3,Q_4) = & -1 - 4\theta_{3,3} Q_3(1 - Q_3) - 3Q_4\nonumber \\
	&  - R + 3R(Q_3 + Q_4 - 1) \nonumber \\
	& + \theta_{3,4} R(4Q_3Q_4 - Q_3 - Q_4).
 \label{V}
\end{align}
By the inequalities of the arithmetic and geometric means, $(Q_3+Q_4) \geq (Q_3 + Q_4)^2 \geq 4Q_3Q_4$, we have 
\begin{equation}
    4Q_3Q_4 - Q_3 - Q_4\leq 0 \ .
\label{C1}
\end{equation} 
Using $Q_3(1-Q_3)\geq 0$, $Q_3 + Q_4 \leq 1$ and (\ref{C1}) in (\ref{V}), we establish that
\begin{equation}
	V(Q_3,Q_4) \leq -1 - R \ .
\end{equation}
Since $R>0$, $V(Q_3,Q_4)<0$ throughout the region to which the system is confined. Thus, according to the Bendixson theorem, this system has no limit cycles. 
Combining this result with the established uniqueness of the steady state~\cite{wanjura_correction} then, by the Poincar\'{e}-Bendixson theorem, the limit set contains only that steady state~\cite{Perko_1996_Diff_Eqn}. Thus, the detailed balance steady state is globally stable for any physical initial condition. \\

\section{Conclusions and future work} 
To conclude, we studied, both theoretically and experimentally, the nature of the detailed-balance steady states into which the non-equilibrium dynamics of granular matter has been found to settle. We have proven that, for any maximum cell order, $\mathcal{C}$, there can only be one detailed-balance steady state. We have also shown rigorously that, if a detailed-balance steady state solution exists up to $\mathcal{C}=7$ then it is the only solution. 

Intriguingly, by solving the steady state equations numerically for $614400$ systems up to $\mathcal{C}=8$, we found that there is always only one physical solution, in which $Q_k$ is real and positive for all $k$. We conjecture that the evolution equations (\ref{eqn:master}) yield only one physical solution for the steady state, which may or may not satisfy detailed balance. This conjecture is supported by clear experimental observations of detailed-balance steady states for systems with $\mathcal{C}>10$~\cite{sun_experimental_2021}. 

Next, we used the theorems of Bendixson and Poincar\'{e}-Bendixson to show that the detailed balance solution of the dynamics of systems with $\mathcal{C}=5$ is globally stable and no periodic orbits exist. This may explain the robustness of such solutions observed experimentally~\cite{sun_experimental_2021}.

It should be noted that, in our discussion, the steady-state rates $p_{i,j}$ and $q_{i,j}$ are constant, but of arbitrary functional form, subject to the condition of detailed balance. However, they need not be constant and evolve during the approach to the steady state. 

Another intriguing implication of our results is the following. We have found that most systems with $\mathcal{C}>5$, settle into steady states that do not necessarily satisfy detailed balance. For example, solving and finding the only solution when $\mathcal{C}=6$,
$p_{3,5} = p_{4,4} = q_{3,3} = q_{3,4} = q_{3,5} = 1$, and $q_{4,4} = 0.5$, we find that not all $\eta_{i,j}=0$, namely, no detailed-balance. 
Yet, the experiments of~\textcite{sun_experimental_2021} reveal that, in a range of quasi-statically cyclically sheared 2D granular systems, the steady states always satisfy detailed-balance. 
This suggests that the cell breaking and merging rates in those experiments were neither arbitrary nor time-independent. Rather, they must have evolved as the granular systems self-organised into values that satisfy detailed balance. 
This seems the most plausible explanation for the emergence of detailed balance in those experiments and could reconcile the discrepancy between those observations and what appears to be a violation of the paradigmatic Klein principle~\cite{klein_principle_1955}.

Rate equations similar to (\ref{eqn:master}) have been used for modelling many evolution processes. In particular, in models of aggregation-fragmentation (AF). Nevertheless, there are some similarities and differences between the granular dynamics we study here and those models. The similarities are that our cell order fractions, $Q_k$, are analogous to aggregate sizes and all models have transition rates. Additionally, many models of AF assume detailed balance, implicitly or explicitly~\cite{AF1,AF2,AF3,AF4,AF5}. 
One minor difference is that the normalisation of our cell order fraction takes into consideration the changing total number of cells, which which gives rise to the last term in our eq. (\ref{eqn:master}).  
There are, however, more significant differences. One is that, unlike in most of AF models, for example, we do not assume the mathematical forms of the steady-state rates, $p_{i,j}$ and $q_{i,j}$, subject to the condition that they satisfy detailed balance. This makes our analysis more general and more applicable than the studies that make such assumptions. 
Another difference is that most studies of AF dynamics simplify the analysis by allowing the size of aggregates to tend to infinity, which is often not physical. In contrast, our analysis applies to arbitrarily finite highest order, $\mathcal{C}$. 
The third difference is quite fundamental. As mentioned, many models of AF processes assume detailed-balanced steady states. While this phenomenon is well established for systems in equilibrium, the current belief in the community is that it cannot be satisfied in out-of-equilibrium steady states. Indeed, our motivation to study steady states of sheared granular system is the surprisingly strong experimental evidence in~\cite{sun_experimental_2021} of persistent detailed-balanced in the steady states of their granular dynamics. 
To the best of our knowledge no such evidence exists for the non-equilibrium systems, to which some AF models presume to apply. This also means that our results apply only to AF systems either in strict equilibrium or where detailed balance has been established experimentally.



\bibliography{Bibliography}

\end{document}